# Towards Investigating Substructures and Role Recognition in Goal Oriented Online Communities

Tanmay Sinha
School of Computer Science
Vellore Institute of Technology
Chennai 600127, India
tanmay.sinha655@gmail.com

Indra Rajasingh
School of Advanced Sciences
Vellore Institute of Technology
Chennai 600127, India
indra.rajasingh@vit.ac.in

*Abstract*—In this paper, we apply social network analytic methods to unveil the structural dynamics of a popular open source goal oriented IRC community, Ubuntu. The primary objective is to track the development of this ever growing community over time using a social network lens and examine the dynamically changing participation patterns of people. Specifically, our research seeks out to investigate answers to the following question: How can the communication dynamics help us in delineating important substructures in the IRC network? This gives an insight into how open source learning communities function internally and what drives the exhibited IRC behavior. By application of a consistent set of social network metrics, we discern factors that affect people's embeddedness in the overall IRC network, their structural influence and importance as discussion initiators or responders. Deciphering these informal connections are crucial for the development of novel strategies to improve communication and foster collaboration between people conversing in the IRC channel, there by stimulating knowledge flow in the network. Our approach reveals a novel network skeleton, that more closely resembles the behavior of participants interacting online. We highlight bottlenecks to effective knowledge dissemination in the IRC, so that focused attention could be provided to communities with peculiar behavioral patterns. Additionally, we explore interesting research directions in augmenting the study of communication dynamics in the IRC.

## I. INTRODUCTION

In recent years, the development of free and open source tools has received great impetus from the educational research community. With the increased use of open source technology, Internet Relay chats(IRC) have become a popular form of synchronous computer mediated communication, among people gradually adopting this technology. Free and open source software holds compelling advantages over proprietary software in terms of improved security, cost, customizability, flexibility and interoperability. This in turn, has given great impetus to the contributors or developers of open source, leading to a proliferation of specialized support forums that are dedicated to help users. Often, these help forums take the form of IRC, where the primary objective is to connect people from different demographics, while facilitating discussion on technical issues and core concepts related with the specific open source product. This kind of informal knowledge sharing takes place outside the confines of a physical institution. It provides unprecedented opportunities for systematically leveraging computing methods to increase the information flow among a big community of people. Because of the extensive reach, power and applicability of open source products, such computing approaches have the potential to be rapidly deployed for assessing their performance and results.

Therefore, we investigate one such extremely popular open source operating system distribution, Ubuntu, which is powered by Linux. Since its first release in 2004, Ubuntu has improved immensely and become a mature desktop distribution. Its true popularity was revealed in 2011, when the hits on Wikimedia pages from computers running Ubuntu had risen from 16 million to 29 million. Currently, traffic analysis reports reveal that this number has exponentially risen to 1100 million [1]. Given this massive scale of adoption, users need to be provided with ample guidance to effectively complete their tasks using the operating system in a timely manner. Because of the inefficiency and complication of specialized and centralized support that can be offered, a peer to peer approach is followed to help community users solve their queries. This peer community in Ubuntu functions via an IRC. Thus, IRC provides a platform for fruitful user engagement. It acts as a channel to ensure effective knowledge dissemination among diverse user groups which collaborate to mutual advantage and enjoyment. As more and more people join the IRC, they provide a multitude of approaches for better conceptual understanding and problem solving, thereby increasing the quality of solutions. This IRC community thus evolves as a goal oriented community, where the primary objective is to solve user queries arising among a motley of interacting interest groups and facilitate discussions among them.

In our work, we are interested in gaining insight into the interesting communication dynamics exhibited in the IRC network, and possibility frontiers for leveraging this information to explore what makes people behave the way they do in online communities. The motivation for this kind of analysis intuitively arises from the hidden mapping between pattern of ties among people in the IRC and their actual interaction patterns. In real world too, people play roles each time they interact and these roles profoundly shape behavior and expectations of interacting participants. Consider person A, who serves as a doctor to person B, a professional colleague to person C, and a family member to person D. The information he shares in his small group is entirely different for person B, C, D in terms of message type, importance and relevance of the message (what), its context (why), quantity (how much), frequency/message time interval (when), place (where) and medium used for communication (how). The contrast is better explained with this example. As a partner, person A will be concerned with building a lasting and loving relationship, while as an employee, he would want to get on with his business of giving quality medical treatments to patients and earning his daily bread in the process. His openness and motives behind communication will vary significantly. This notion of role influenced behavior can also be generalized to distributed settings where people may be interacting online (remotely), a prominent example of which is an IRC. Therefore, labeling people by roles can give us an evaluation measure of the goals and standards of achievement that people need to reach and an expected behavior that they are supposed to exhibit. We can informally discover the nature of social interactions, if these recognized roles are mapped to conversations.

This paper is organized as follows: In Section II, we describe the motivation that drives our research. Section III deals with the related work relevant to this domain Section IV describes our dataset chosen for experimentation with justification for the same. This section also highlights the social network extraction out of the Ubuntu IRC. In Section V, we characterize this IRC network by sub structure identification. We discuss some interesting research directions in Section VI. To cut a long story short, we conclude by talking about future work and the implications of this research in Section VII and VIII.

## II. MOTIVATION

Among other forms of computer mediated communication, it is interesting to examine IRC's because "people who are located in geographically distant locales, who are of different national and linguistic backgrounds, and who might otherwise never come into contact, can engage in real-time interactions that resemble the immediacy of in-person face-to-face encounters" [45]. To study the Ubuntu IRC community in a perspective influenced by learning analytics, we choose a social network lens. Social network analysis(SNA) methods form the basis of identifying possible roles that people play in their relationships with one another, which further leads to contrasting social influences. Adopting such an approach helps to decipher informal connections between people conversing in an IRC channel, which may not be otherwise visible. Literature suggests that "organizational members benefit from external network connections because they gain access to new information, expertise and ideas not available locally, and can interact informally, free from the constraints of hierarchy and local rules [2]". For understanding the social structure hidden behind the IRC that facilitates sharing of skills, it is necessary to focus on actors and relations, rather than adopting a more conventional data driven approach that stresses on actors and attributes. This difference in emphasis is consequential for the choices we make in our research. Before proceeding further with our analysis, it is important to understand real world implications of achieving clarity about communication dynamics by exploiting social connectivity patterns. Our analysis will:

1. Enable the development of a rigorous model for information flow that captures real world notion of abstract roles more closely. This can be used for:

    a. Implementing computer supported collaborative learning tools to assist and assess knowledge transfer between people who participate in online communities, such as in the work of Trausan-Matu et al. [30]

    b. Filtering IRC messages to split the people conversing into different groups for greater efficiency

2. Immensely benefit open source developers and the learning community by prioritizing and capturing the change in information, ideas and views relatively rapidly. It is these ideas that shape network structures, which in turn controls where the information can spread.

3. Allow better inference for the possible reasons for abnormal/unexpected behaviour in any part of the network more clearly, while helping developers to narrow down focus on groups showing diverse community involvement. For example, in the context of Ubuntu IRC,

    a. Huge quantity of unexpected information flow leading to increasing traffic on particular links, can enable us to extract and leverage frequently occurring linguistic patterns that can be automated through bots or cognitive tutors, to motivate other groups of people

    b. On the other hand, extremely low communication traffic can help us to devise schemes to offer specialized support to certain individuals.

## III. RELATED WORK

In this section we outline the perspectives on communication dynamics that have been explored so far in the emerging literature on online communities. Prior work draws from and contributes to various bodies of literature. Our literature survey is grounded in the use of social network methods for studying online communities, especially IRC's. Uthus et al. [32] identified social phenomenon detection and topic detection as prominent research areas for multi participant chat analysis, among other relevant directions like chat preprocessing, chat room feature processing, thread disentanglement, user profiling, message attribute identification and automatic summarization. The research in applying social network methods to analyze dynamically changing chat interactions received impetus through the development of IRC bots. Mutton et al. [34] implemented an IRC bot to automatically infer dynamically changing social network out of the IRC interactions by applying methods like direct addressing of users, temporal proximity and density. However, more focus was on better visualization of the evolving network, and not on determining the larger impact of these inferences for improving learning within the IRC. Similar work was carried out by Camptepe et al. [35], but with a modified objective of community detection. Tuulos et al.'s [33] primer work on combining topic models and social networks, by filtering non important parts of a discussion using social relations between chatters, opened up a plethora of rich literature on web graph analysis for multi participant chat. The transformation of social language processing into social language network analysis (SLNA) has provided a means to address questions such as the role of social relationships and group coordination in communities. This perspective aligns well with the sociolinguistic literature, because it involves understanding linguistic variation with respect to social position [42].

In the spirit of these types of historical developments, Scholand et al. [43] in their work aptly highlighted the effectiveness of applying SLNA to a real world knowledge intensive collaborative work communication corpus. Their methodology demonstrated explicit important components of organizational functioning, such as information exchange and evaluation (a function of perceived expertise) and social support. Paolillo et al. [45] studied the relation between network tie strength and the frequency of the use of different linguistic features, by applying factor analysis methods. The results showed a complex distribution of features like obscenity, use of native language and abbreviations across core and peripheral groups. Related to this, Ștefan et al. [46] computed a one to one mapping between social factors and semantic utterances of students participating in computer supported collaborative learning chat forum. For every node pair (i,j), all the marks of the utterances exchanged between them were evaluated semantically using length, frequency, branching factors, likelihood between the terms used in the current utterance and the whole document etc. Such learning analytics was used for giving grades to individual students based on their discourse quality. However, our focus is more at a macro level, than a micro level view of the network. Vladimir et al. [40] studied the hierarchical organization of the Ubuntu IRC social network with a mapping to the most frequently used message types and the amount of emotional arousal in them. IRC message annotation was performed using the Senti Strength classifier, Linguistic Inquiry and Word Count dictionary(LIWC) and Affective Norm for English Words(ANEW) based classifier, designed for measuring word usage in psychologically meaningful categories.

The findings showed that message types were heterogeneously distributed among users and despite content based linking in the channel, users not in the central network core tended to develop a social network of their own. In an attempt to better infer social behavior from community networks, other interesting research directions involve detection of deceptive chat in online communication [36][37], studying emotional persistence [38] and factors of social capital like etiquette, empathy and trust [39]. Though some of these works have examined the linguistic behavior in correlation to network structures like tie strength and centrality, none of these works specifically explore communication dynamics using social network analysis concepts we have systematically applied, and link abstract roles of people in goal driven communities to nature of their conversations.

## IV. STUDY CONTEXT

### A. Dataset & Characteristics

We chose the Ubuntu IRC Chat Corpus (UCC) for our purpose of social network analysis. To build this corpus, we scraped the archived chat logs from Ubuntu's Internet Relay Chats' technical support channel. Specifically, we selected the "#ubuntu_beginner" support channel of Ubuntu as our forum for experimentation. The primary focus was to address the growing popularity of open source IRC forums in the past few years because of the advancement in digital media, rapid deployment of open source systems, increasing user and an evolving developer community. Therefore, these logs spanned a period of 2011 and 2012. These chat records were organized on a daily basis for each month. Each log file had a timestamp, username and his conversation. In total, our dataset consisted of 166581 IRC messages from 3470 users.

The "#ubuntu_beginner" channel is basically dedicated to new comers who seek a multitude of fundamental answers about the operating system distribution, its usage, features and development process. We heavily debated the source of data for this experiment. Finally, we chose the particular channel because the discussions and real time problem solving in this beginner channel are crucial for drawing in more people into the open source community. The prompt support offered from the community members forms a basis for active participation of people, healthy conversations and future possibilities for people to join together and contribute on open source research projects. Also, among all the other primary IRC channels of Ubuntu, the chosen channel was expected to exhibit a lot of diverse community involvement, in terms of contrasting participation behavior of: (a)beginners or question askers and, (b)repliers or people belonging to the expertise network. Therefore, we were interested in studying the sub structures and roles in this particular IRC channel.

The lack of structure or subdivision into threads and sub-threads was a preprocessing disadvantage with this dataset and had the potential of distorting our analysis results. This difficulty in dealing with informal characteristic of chat grammar as compared to traditional text has been highlighted in existing literature on analysis of multi participant chat [27][28][29]. However, our IRC dataset still provided ample opportunities for exploratory data analysis and knowledge discovery via social network methods. The textual IRC logs provided information on social interaction as well as linguistic usage which was of interest. The advantage of using this dataset was that it contained technical discussions, which allowed for research that is applicable to other technical domains(e.g - business, online courses, collaborative learning, military command and control). The presence of less social chat than a non-technical chat channel, helped us to filter interesting on topic discussions going in the IRC for our purpose of analysis. It was more reflective of knowledge building in the IRC discussions.

## B. Extracting a social network from the Ubuntu IRC

Prior works have investigated formation of social networks in multi participant chat analysis using approaches like temporal coherence, reply structure, conversation proximity, and word context [3][4][5][6]. In our work, the last approach seemed apt for building a social network out of the Ubuntu IRC data. Because IRC's can have multiple, asynchronous and interwoven conversations going on simultaneously among diverse user groups, it is a preferred IRC etiquette that users refer each other via their usernames while addressing conversations. Such forms of targeted messages enable easier understanding for the recipient of the message and eliminate confusion that can build up during peak usage of the channel. We observed that this norm was being followed in majority of the cases; in some cases, at least for the first set of group conversations. This direct addressing allowed us to establish ties between people using the IRC.

Therefore, to try to maximize the effectiveness of our work, the network formation consisted of outwards links from message senders to the recipient names as mentioned in the message. Having created a prior list of all people using the Ubuntu IRC, we searched for these names (including regular expression patterns) in the conversations of participants to check whether they were referred or not. Multiple references led to a stronger tie between nodes of the IRC (people). Thus, the directed weighted network consisted of IRC users as nodes, while the strength of connections or edge weights of the social network corresponded to communication frequency. The example below indicates a sample conversation between users. According to our network formation procedure, we shall have a directed link from "mdz" to "lifeless", and "fabbione" to "mdz".

*[08:43] <mdz>* **lifeless:** *ok, it sounds like you're agreeing with me, then*

*[08:45] <fabbione>* **mdz:** *i think we could import the old comments via rsync, but from there we need to go via email. I think it is easier than caching the status on each bug and than import bits here and there*

## V. CHARACTERIZING UBUNTU IRC NETWORK

To investigate the pathways through which knowledge transfer takes place in the IRC, it was critical that we first explored the notions of sub-structure identification in the IRC, which could be further generalized. Therefore, our research methodology comprised of the following coherent steps:
1) Inferring the communication dynamics of the IRC network by building a high level skeleton of the network.
2) Determining social hierarchy (power, position and importance of people) by node/edge centric analysis, specifically by analysis of cliques, blocks and lambda sets based on distributed information flow.

We utilized software packages like Ucinet [7], Gephi [50] and Pajek [51] for some of our experiments. For our purpose of analysis, we considered the social network for the #ubuntu_beginner IRC channel for the year 2011 and 2012 separately to compare and contrast the findings. The 2011 social network consisted of 2383 nodes and 9354 edges, while the 2012 social network consisted of 1087 nodes and 2699 edges. Figure 1 and Figure 2 show the IRC network for 2011 and 2012, where the size and color of nodes depicts the HITS (Hypertext induced topic selection) authority and hub scores respectively [22].

In the IRC network, people with high authority scores are those who are active repliers (high information content in the node), while people with high hub scores represent those people who interact and engage with a lot of other people in discussions (quality of nodes' links). Both these figures portray a fairly low and proportionate distribution of the authority and hub scores, even in denser clusters. This indicates lack of effective and healthy discussions in well connected user groups of the IRC. In some of these clusters, we also observe that there are people with very high authority scores, but low hub scores. Thus, we can infer the fact that, the expertise nodes are primarily involved in giving short and objective reply to people's queries, rather than engaging in detailed discussions with the group of beginners connected via common problems. Social network relationships tend to be dominated by weak ties, because participation in the IRC is transient and constantly changing.

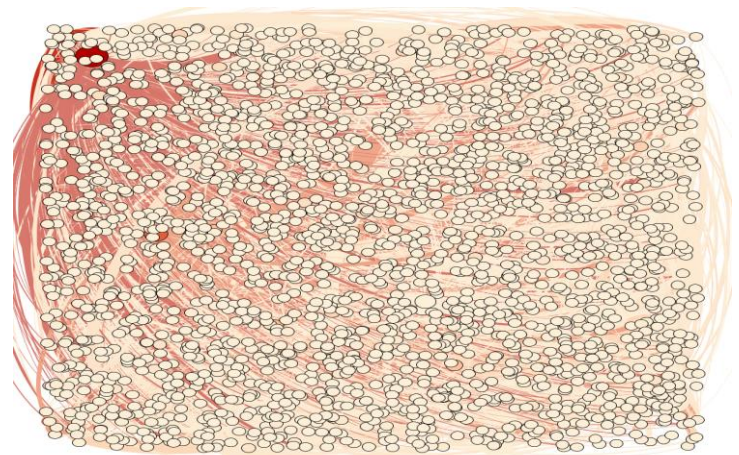

Figure 1. Social network for #Ubuntu_beginner IRC- 2011

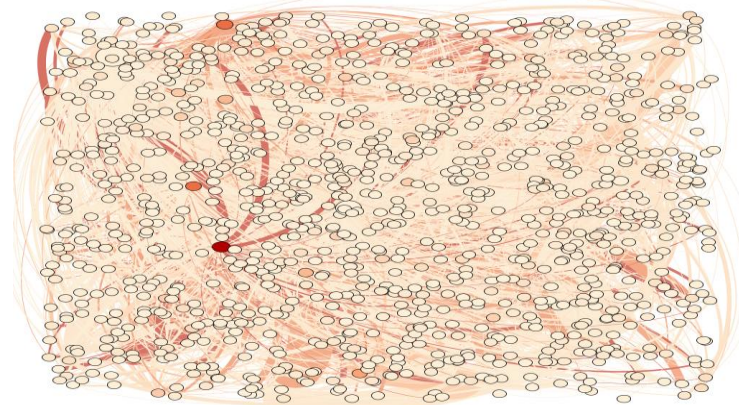

Figure 2. Social network for #Ubuntu_beginner IRC- 2012

## A. Examining the network skeleton

Formally, Bow Tie analysis has been one of the principle ways of investigating the skeleton of networked data [47][48]. The structure divides a graph into 6 main components: In, Out, Scc (Strongly connected component), Tubes, Tendrils and Others (disconnected network nodes). If G = (V, A) is a digraph and S is a strongly connected component of G, the bow-tie decomposition of G with respect to S consists of the sets of nodes as described in Table 1.

TABLE I. BOW-TIE STRUCTURE DESCRIPTION

| Graph Component | Set of nodes |
| --- | --- |
| SCC | S |
| IN | {v ∈ V − S | S is reachable from v} |
| OUT | {v ∈ V − S | v is reachable from S } |
| TUBES | {v ∈ V − S − IN − OUT |v is reachable from IN and OUT is reachable from v } |
| INTENDRILS | {v ∈ V − S |v is reachable from IN and OUT is not reachable from v} |
| OUTTENDRILS | {v ∈ V − S |v is not reachable from IN and OUT is reachable from v} |
| OTHERS | V − S − IN − OUT − TUBES − INTENDRILS − OUTTENDRILS |

However, we notice a very different high level view of the Ubuntu IRC network dynamics after examining the pattern of ties that shape the network. We divide the social network graph into primarily four major components - A,B,C,D. Each component consists of a disjoint set of nodes. Component A is strongly connected. As depicted in Figure 3, the 2012 network shows an almost equal distribution of people in the B and C component, while the 2011 network shows that majority of the people belong to the D component, with a very small C component. In both cases, the proportion of people in the A component is small. Each network component will encompass the interactions of beginner and expert people. Except people in the A component, sparse question answer dynamics are expected to characterize the B, C, D components. Quantifying the connectivity patterns within and outside these components might clearly bring out some fascinating differences between the 2011 and 2012 Ubuntu IRC network structure.

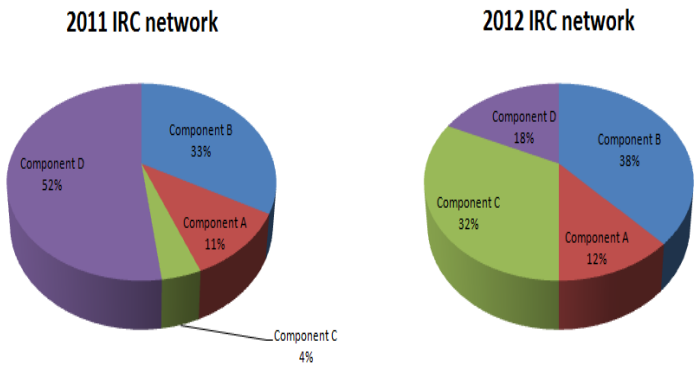

Figure 3. Network skeleton for 2011 and 2012 Ubuntu IRC network

Figure 4 and 5 portray our novel communication dynamics model for the IRC networks. As a sharp contrast to the Bow Tie network structure, this model is less restrictive in defining who can communicate with whom. There is a much more flexibility in the way people interact with each other. This network skeleton is more close to real life behavior of people in the online communities, and depicts that there is a good knowledge flow in the network. The interactions captured here are unlike the Bow tie model, which in the context of the Ubuntu IRC, would be very likely to provide wrong interpretations. The directionality of links would cause the communication pattern in the Bow tie to necessarily move only from the IN component towards the OUT component. However, in online communities it is difficult to get such mutually exclusive subsets of people who only receive or only send messages (only ask or only answer questions).

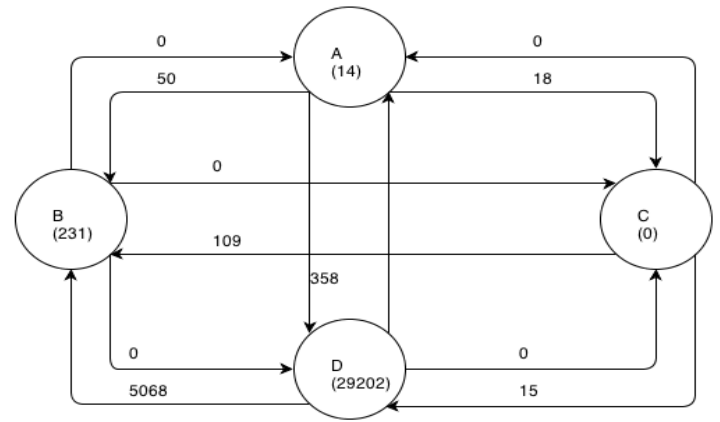

Figure 4. Network skeleton representing number of communication links among sets of disjoint nodes for 2011 Ubuntu IRC network

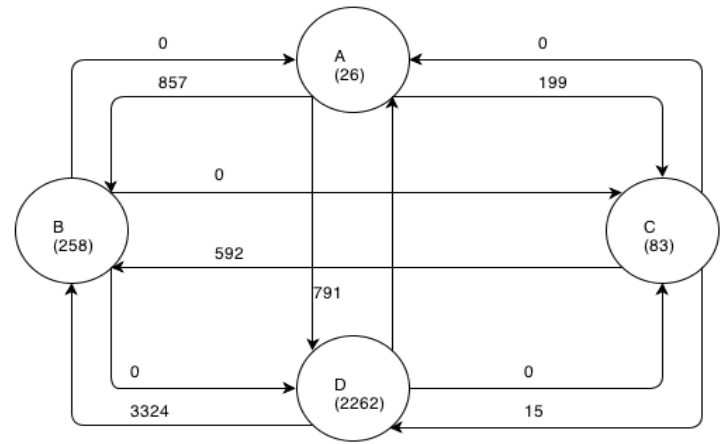

Figure 5. Network skeleton representing number of communication links among sets of disjoint nodes for 2012 Ubuntu IRC network

Some interesting conclusions that follow from this investigation are:

1. In 2011, component C has 0 links within them. However, there are outward links to all other components A,B and D. So, it can be inferred that this community either has all expert members who do not find much need to talk to each other and therefore they help others. Or, it consists of all beginners, who do not have enough knowledge to discuss anything among themselves and therefore they approach people in other components.
2. In 2012, we can observe the presence of 83 links in component D. This finding complements the increase in proportion of the people belonging to component D in this year. There is a mixture of experts and beginners who are discussion among themselves, apart from contacting people in other components.
3. In both the years, we observe a similar pattern of people in components A and C not having much communication links between them. However, they have outward links to all other components. These represent communities of people who have greater relative expertise and help people voluntarily, rather than reciprocating.
4. Because people often offer help voluntarily, beginners questions are not often reciprocated by answers from experts. Therefore, only a small proportion of experts who reply to beginner's questions actually get connected with them. We corroborate this notion in the discovery of substructures in the IRC network in the later section.
5. In both the years, component D covers major proportion of the links among all other components, despite comprising low proportion of people in 2012.
6. People in component A generally support or communicate more often with components having greater communication frequency amongst themselves. They exhibit a kind of preferential attachment, a well studied social network phenomena.

*B. Substructure Identification*

The connection topology of nodes in a social network can be seen from the top down or bottom up perspective. It is consequential for predicting both the opportunities and constraints facing groups and actors, as well as predicting the evolution of the graph itself. Divisions of actors into groups and sub-structures can be a very important aspect of social structure. It can be important in understanding how the network as a whole is likely to behave. Knowing how an individual is embedded in the structure of groups within a net may also be critical to understanding his/her behavior. Such differences in the ways that individuals are embedded in the structure of groups within in a network, can have profound consequences for the ways that these actors see their "society", and the behaviors that they are likely to practice.

*1) Cliques:* Clique is a fundamental graph theoretic concept, formally defined as a maximal complete subgraph. The notion of a clique can be seen as formalizing the notion of a primary group. Our analysis revealed that there were 688 cliques of size greater than 3 in the 2011 network, while only 44 cliques of size greater than 3 in the 2012 network. But on the contrary, average clique size was almost 8 times in the 2012 network (1087/44), as compared to the 2011 network(2383/688). This indicated that smaller and denser clusters changed to larger and denser clusters with increase in communication frequency for the 2012 network. To further investigate these sharply contrasting patterns, we were further interested in seeing the extent to which these sub-structures overlapped, and which people were most central and most isolated from the cliques.

In the 2011 network, finding actor by actor clique co-membership matrix immediately brought two prominent people "holstein" and "bioterror" into limelight. These two people in the IRC had a maximum number of 137 cliques in common. Figure 6 depicts the ego networks (direct one degree ties and relationships among them), where the nodes are colored by the number of inlinks (indegree centrality). The dense ego networks with disproportionately connected alters reveal that these people hold a very strong structural position in the IRC network. The clique participation scores or the proportion of clique members that these nodes were adjacent to, further strengthened our quantitative findings. However, when we dived into the 2012 IRC network, we found a fairly proportionate clique by clique actor co-membership. All the 44 cliques discovered had only 1-4 members in common. The maximum size clique also comprised of only 3 members, as a contrast to the 2011 IRC that had a maximum clique size of 9. In 2012, "holstein" and "stslaint" had a maximum clique overlap of 10 cliques. To examine how had the persistence of this interesting IRC user, "holstein" changed with time, we looked at his ego network in 2012. It can be clearly observed in Figure 7, that the overall size and density of the ego network has reduced.

Multi participant, dynamically changing and primarily question-answer driven online communities such as IRC's are expected to resemble more of a clique structure, rather than a star network. This is because, in the Ubuntu IRC network, people have questions about their particular domains of interest (parallel programming, web development etc.) and would want to seek specific answers, as to how open source technologies can aid in problem solving. So, the network would more likely be clustered at different places, with people getting more closely and intensely tied to one another with similar interests, than they are to other members of the network, eventually leading to clique formation. In contrast, a star network would be more applicable to communities such as those formed in massive open online courses(MOOC's), where there are limited number of expert students who participate in discussion questions with others, within a particular domain. This was the core idea behind investigating cliques.

However, if everyone could communicate with everyone in a clique, then we required a solution for determining leaders in the clique. Intuitively, if there was a node that connected a particular clique to a different clique or network substructure, we would want to focus on those nodes. This formed the motivation for examining Blocks and Cutpoints in the IRC network.

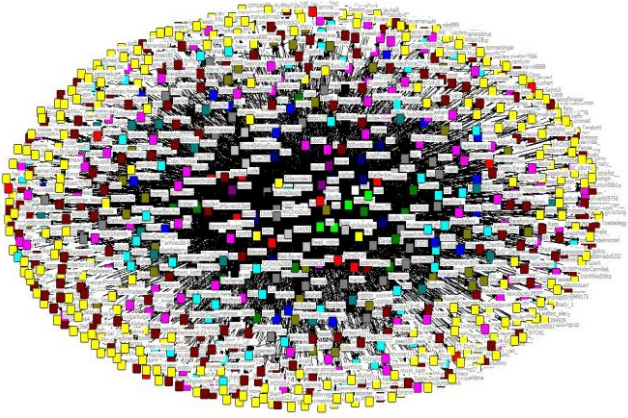

Figure 6. Ego network for IRC user "holstein", 2011

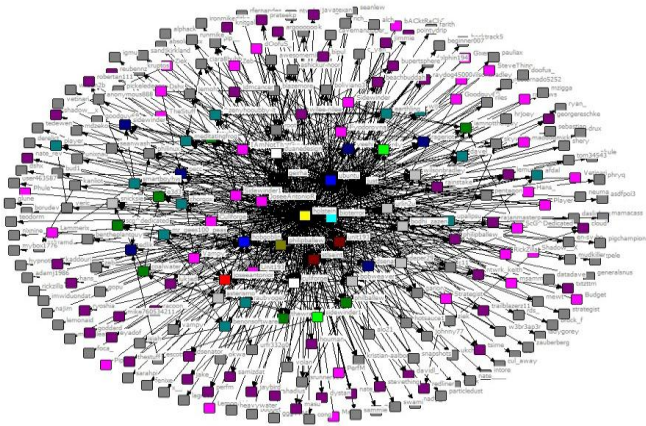

Figure 7. Ego network for IRC user "holstein", 2012

*2) Blocks & Cutpoints:* Next, we examined the key "weak" spots in the social network graph. The question we asked was, if a node were removed, would the structure become divided into un-connected parts? And what interesting characteristics could we decipher from the blocks or divisions into which cut-points divide the IRC network? This could help us identify such nodes or cutpoints, who were important people in the IRC network and facilitated discussion among disconnected groups. We found out that the 2011 IRC network consisted of 196 articulation points and 867 blocks, with one peculiar block comprising of 1503 IRC participants. This block had both "holstein" and "bioterror" as members, indicating their strong betweenness centrality and further strengthening intuitive notions of their importance in facilitating IRC discussions. The 2012 IRC network on the other hand had 495 blocks and 105 cut points, with only one large block having 588 people.

Both the 2011 and 2012 network exhibited the consistent formation of a very large block. One of the reasons could be the fact that, this big group of people are engaged in solving similar, but peculiar queries/problems that are highly important and require immediate attention. Because of the disinterestedness of others in these discussions, rest of the blocks are very small in size. This indicates that beginners are not drawn much into the open source community and very often, these beginners are connected only to only a small proportion of the IRC expertise network. From a learning perspective, we can very well point out that effective knowledge dissemination will not happen in the network, because of fairly high localization. Sparsity of the overall network (avg density = 0.0065±0.35 for the 2011 and 2012 network) is evident from these findings. Thus, for increasing the interaction between beginners and expert people, we need to focus on key actors and locate parts in the IRC graph that are vulnerable. Having this knowledge, we can provide incentives for these expertise nodes to continue sharing their knowledge and helping different subgroups to collaboratively work on open source problems.

At this junction, it is also important to note that most of the newbies in the small blocks engage in a short conversations, sometimes even receiving no replies. The reasons for this lack of attention might be improper question formation or the question being out of context (not relevant to IRC channel scope), too specific/localized (answers to which are not known), uncivil/impolite (not conforming to the IRC etiquettes) or asked at the wrong time(when not many IRC users are online). We highlight few such cases from the 2011 IRC network in Table 2.

TABLE II. PROBABLE REASONS FOR SHORT CONVERSATIONS

| Category | Text |
|---|---|
| Uncivil Behavior | *"this is a support channel and we are all volunteers to help support users;**venting** at us won't be particularly useful and will only demotivate us"* |
| Improper Question formation | *"I am running Ubuntu 10.04 and **for some reason** when I clear recent documents, it keeps reappearing"* |
| Out of Context | *"Hi there, who knows whether **ebook readers and kindles** are able to process (x)html <table>?"* |
| Too Localized | *"Been trying to install Lubuntu on an, **old system(P3 500)** with problems..any helpers out there"* |

*3) Lambda Sets:* While the previous approach focuses on actors in a social network, sometimes we might also alternatively be interested in certain connections in the graph which, if removed, would result in a disconnected structure. If l(a,b) represents the edge-connectivity of two vertices a and b from a graph G(V,E), then a subset S is a lambda set if it is the maximal set with the property that, for all a,b,c ∈S and d ∈ V-S, l(a,b) > l(c,d). The motivation for examining the Lambda sets in the Ubuntu IRC network thus arises because of the importance of relationships between people, something that is fundamental to the study of social networks. The Lambda set approach ranks each of the relationships in the network in terms of importance by evaluating how much of the flow among actors in the net go through each link. It then identifies sets of relationships which, if disconnected, would most greatly disrupt the flow among all of the actors.

Interestingly, in our Ubuntu IRC network for 2011 as shown in Table 3, we see that the maximum information flow takes place mainly via some of the most important people in the substructures identified or via nodes holding important structural position in the network. The reasons for this preferential attachment phenomenon may be node popularity ("rich get richer") in the IRC, or increasing node quality ("good get better"). Thus, Lambda sets highlight points at which the fabric of social connection in the Ubuntu IRC is most vulnerable to disruption. There are two advantages of examining the lambda sets: (a) people in the IRC who maintain multiple, independent, paths of communication are more likely to share linguistic practices and norms of the IRC, than those with few paths, if there is a high likelihood of edge destruction, and (b) Minor problems will not necessarily escalate into group fission, because bad feeling between any pair of members will not cut the subset in half: all the other members in the lambda set will continue to be connected to each other despite the missing link. Table 3 highlights some of the top communication links with maximum information flow.

TABLE III. LAMBDA SETS

| Year | Top Links with maximum information flow |
| --- | --- |
| 2011 | ubuntu-holstein, stslaint-holstein, ubuntu-stslaint, bioterror-stslaint, holstein-bioterror, ubuntu-bioterror |
| 2012 | ubuntu-holstein, just-holstein, stslaint-holstein, ubuntu-stslaint, geirha-holstein |

## VI. RESEARCH DIRECTIONS

In this section, we discuss some of the interesting research directions that can aid in understanding goal directed communities such as the IRC with greater clarity from a learning perspective.

### A. Role Recognition

Though the graph theoretic approaches described in Section 5 clarify the obvious notions of what constitutes a network substructure, they do not address the group selection issue. The optimal set of players may not be the same as the set of players that are individually optimal, i.e, people in the Ubuntu IRC network may be redundant with respect to their liaising role (they might be connecting the same third parties to each other). Therefore, it would be more effective to group together actors who are the most similar, describe what makes them similar and, to describe what makes them different, as a category, from members of other categories. The interactions of people involve behaviors associated with defined status and particular roles. These status and roles help to pattern our social interactions and provide predictability [14]. The role categories are defined in terms of similarities and regularities of the patterns of relations among actors, rather than attributes of actors. Three particular definitions of equivalence have been particularly useful in applying graph theory to the understanding of social roles and structural positions that can facilitate effective knowledge dissemination in communities.

Actors that are structurally equivalent [8] are in identical positions in the structure of the network sociogram. Whatever opportunities and constraints operate on one member of the IRC are also present for the others. Having this abstract idea, we can afford to replace IRC members during their absence in the community without significantly hampering the flow of ongoing discussions. However, exact structural equivalence is rare (particularly in the Ubuntu IRC network). Thus, it would be more appropriate to find substitutable substructures rather than substitutable individuals. Else, two beginners, despite similar patterns of interaction with different people in the IRC community, will not necessarily seen as playing the same role. This emphasis on the groupings and relationships is realized through the idea of automorphic equivalence [9], which resolves the fundamental difficulty with using structural equivalence as a formalization of the role concept. Automorphic equivalence is more generalizable to real world networks like IRC's which can exhibit structural replication. In the Ubuntu IRC, automorphic equivalence classes are groupings who's members would remain at the same distance from all other actors if they were swapped, and, members of other classes were also swapped. The reason why this is significant from a learning perspective is that many different people in the IRC discussions can be actively involved in helping newcomers to solve their queries, rather than a fixed subset of people.

Nevertheless while we can effectively detect individuals in the IRC network who have the same network positions or locations with respect to other individual actors, it is difficult to describe social roles unless we find out people who have the same kinds of relationships with some members of other sets of actors. This notion is captured by regular equivalence sets [10], where two nodes are said to be regularly equivalent if they have the same profile of ties with members of other sets of actors that are also regularly equivalent.

Prior work has concentrated on the importance of roles in education [15], healthcare [16], human-computer interactions and games [17][18], social interactions such as negotiating salary/hiring interviews/conducting speed-dating conversations [19], information retrieval [20] and semantic segmentation of data [21]. One of the recent works by Strzalkowski et al. [31] described primer work on discovering social phenomena behaviors such as topic control, task control, and leadership. They considered a two-tiered approach, with the higher-level tier consisting of social roles such as leadership and group cohesion, which builds upon a mid-level tier consisting of social behaviors such as topic and task control. However, there is lack of formalism in connecting role recognition to multi participant chat analysis.

Therefore, we made a first step for diving into the individual components A, B, C, D in the network skeleton and see whether we could quantify the notion of roles, associated chaos and redundancy to some extent. Firstly, using REGE algorithm [53], we compute pair wise regular equivalence between people in the IRC. Two actors are regularly equivalent if they are equally related to equivalent others. REGE is an iterative algorithm where, within each iteration, a search is implemented to optimize a matching function. The matching function between vertices i and j is based upon the following. For each k in i's neighborhood, we search for an m in j's neighborhood of similar value(quantitatively). A measure of similar values is based upon the absolute difference of magnitudes of ties. This measure is then weighted by the degree of equivalence between k and m at the previous iteration. It is this match that is optimized. This is summed for all members of i's neighborhood over all relations and normalized to provide the current iteration's measure of equivalence between i and j. The procedure is repeated for all pairs of vertices for a fixed number of iterations. The result of this iterative procedure is a symmetric similarity matrix which provides a measure of regular equivalence. We choose 3 step neighborhoods for our implementation purpose.

Next, we grouped and ordered the regular equivalence matrix according to the components discovered in our IRC network skeleton. We chose a regular equivalence threshold of 0.5, and calculated the proportion of people having greater than certain percentage of their connections with high regular equivalence (> 0.5). The implications from a learning perspective are outlined in Table 4. As an enhancement, it would be interesting to tune these thresholds to see, at what thresholds do we get the most clear picture of real world IRC behavior.

TABLE IV.   PRIMER ROLE RECOGNITION EXPERIMENTS

| % of connections (ties) with high regular equivalence | % of people | Probable characteristics |
|---|---|---|
| Case 1: More (>30%) | More | 1 big role, Restricted opportunities, Most redundancy, Least chaos |
| Case 2: More (>30%) | Less | Different roles, Greater chaos than case 1, Lesser redundancy than case 1 |
| Case 3: Less (<30%) | More | Many different roles, Least redundancy, Most chaos |
| Case 4: Less (<30%) | Less | Many different roles, Greater redundancy than case 3, Lesser chaos than case 3 |

Figure 8 depicts the graphical representation of our experimental results after setting high regular equivalence to be greater than 0.5. In 2011, components A and B exhibit Case (i) behavior, while components C and D exhibit Case (iii) behavior. Since, these attitudes lie on the extremes, it would be good to shuffle people within these 2 clusters, so that average overall chaos within each component is minimized, while minimal redundancy is retained. In 2012, components A, C, D exhibit Case (iii) behavior, while component B exhibits Case (iv) behavior. To minimize overall average chaos in all components, people in component B would have to be shuffled across the other 3 components. Firstly, we would have to quantify the notions of chaos and redundancy using social positioning of people in the IRC. Intuitively, there would be chaos and confusion in a network if people had more than enough options to contact and consult others. For example, we could use extremely high betweenness centrality measures as one of the metrics for chaos. Next, an optimization approach to this problem could be adopted, wherein we would have to balance the tradeoff between redundancy and chaos. At each step, we would have to compute the average redundancy and chaos across all components, and accept the new solution, if it was better than the previous iteration. Approaches such as Hill climbing, Simulated annealing and Genetic algorithms might be handy.

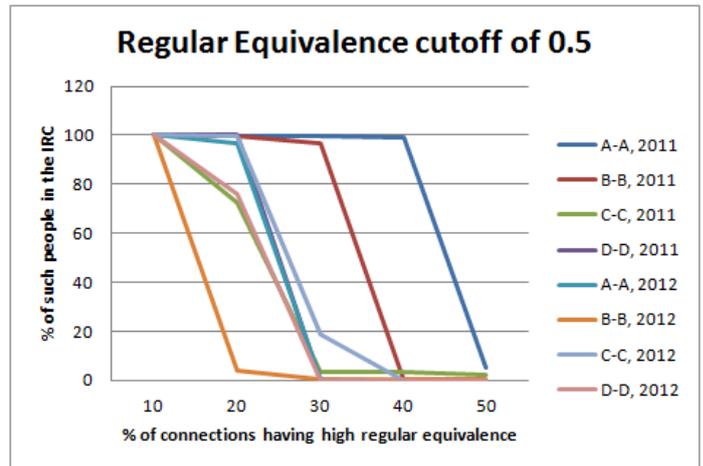

Figure 8.  Experimental results on the IRC network at a regular equivalence cutoff of 0.5

B. *Aligning Social Roles with Linguistic Behavior*

Marshall et al. [44] in his work on sociolinguistics asserted that "speakers use linguistic variation to actively construct and negotiate personal and group identities". Therefore, viewing difference in social roles from the perspective of changed course of action and line of thought across these roles might be fascinating. After a systematic application of quantifiable social network metrics, the social network perspective could be connected to a linguistic lens, to view the interaction patterns and stylistic aspects of conversations are different and make specific groups of IRC participants distinct in the networked learning community. Text mining and topic modeling methods can be used to understand the themes behind actual information content to identify behavioral patterns in information shared with different people. This is because very often is such online communities, recurring usage of language patterns develop as people regularly return to the IRC to resume and maintain the relationship they initiated.

## C. Topic Analysis with varied objectives

Applying topic analysis in detecting which topics are distinct IRC users interested in conversing would be another research direction. This can be helpful in determining if an IRC conversation is on topic or not in different contexts such as computer supported collaborative learning, matching users to appropriate chat rooms based on dynamic topic analysis and its mapping with user's likes and dislikes, optimizing chat queries, tracking topic changes within the channel etc. The machine learning approaches that have been leveraged for studying linguistic behavior and topic models [11][12][49] are based on the assumption that the corpus is a static set of documents written in a fairly comprehensible and accurate manner. However, adopting a more specific data driven approach, suited to the language used in IRC's would do more justice in deciphering the significance of relationships among words and discussion themes of people conversing.

## D. Predicting Survival of people in the IRC

Using the linguistic behavior of people to predict their survival in the IRC is an equally fascinating and closely connected pathway that can be taken up. The issue of persistence in IRC communities like Ubuntu is of prime importance, because the IRC would not serve its purpose of promoting collaboration unless users stay for long periods on the online platform and engage in fruitful discussions. The ultimate question is whether we can track and predict beforehand, which people are going to leave the IRC community based on their conversation patterns. The motivation for such an analysis comes from an interesting recent work by Danescu et al. [41], that stresses on linguistic change as a potential indicator to predict people's activity in popular beer communities. Also, student's forum behavior and their social positioning [54] within the discussion forums, have been used to successfully and effectively predict their survival in Massive Open Online courses [55].

## E. Studying evolution of IRC etiquettes

A slightly different perspective would be to track the evolution of norms or informal etiquettes in the Ubuntu IRC community, such that newbies can merge well with people having expert profiles. In virtual communities such as Ubuntu, where anonymous people of diverse interests interact dynamically, the common code of behavior helps to sustain cooperation. Prior work has already shown the positive impact of group norms in eliciting response in the goal driven Ubuntu IRC community [13].

## F. Improving Social Network formation

Adopting automatic threading approaches to effectively handle chat discussions, as dealt in prior work [23][24][25][26] is a brilliant way to make the IRC data more structured, before creating a social network out of it. Applying such methodologies in open source online communities might overcome the difficulties of dealing with entangled conversation threads and increase the accuracy of social network formation.

## VII. FUTURE INSIGHTS

In general, understanding information flow can play a crucial role in a range of other behavioral phenomena too, including dissemination of information (viral marketing campaigns), adoption of political viewpoints (blogs and propagations), technologies and products (product penetration, recommendation systems). Though we have currently applied the above graph theoretic approaches to #ubuntu beginner IRC community, it is applicable to other goal oriented communities as well. There are some interesting future directions, that can follow up from this preliminary investigation of the #ubuntu beginner IRC community. Some enhancements that would be interesting to pursue are:

1. Comparative analysis of the chosen method of social network formation in the IRC, with other approaches like reply structure of people and time coherence. Such a rigorous methodology would enhance the social network interpretation heuristics.
2. Contrasting the differences in results using a larger chunk of the Ubuntu IRC dataset, with logs spanning from the beginning of the Ubuntu IRC community development (2004) till 2012. Because of huge underlying community of people who participate in Ubuntu IRC channels, effective parallel methods for dealing with this high dimensional data need to be applied.
3. Examining the proportion of communication among people that just passes through important links in the IRC network, versus the information that is solely shared between the two prominent vertices (people) who share the link. This will help us to more clearly understand whether (a)Discussions on queries of newbies in the #Ubuntu_beginner IRC channel are facilitated via these links, or (b)Newbies are left out of discussions and problem solving, because traffic on important links that corresponds to important IRC users, outweighs the traffic due to multiple queries from new users of the channel.
4. Investigating brokerage roles played by members of the IRC, such as coordinator, consultant, gatekeeper, representative and liaison [52].

## VIII. CONCLUSION

In our work, we made a primer investigation into communication dynamics of the #ubuntu beginner IRC channel. We were able to provide a new perspective to examine the network skeleton. Identifying sub structures in the IRC network helped us to generalize the pattern of relationships between people and enabled us to discover some possible bottlenecks to knowledge flow in the network. The fundamental objective was to understand why some people behave in distinctly different ways in their IRC conversations, despite being "birds of the same feather flocking together" and vice versa. Gaining such useful insights could enable open source developers to design novel strategies for improving the IRC learning environment and making it a better networked learning community. Such artificial intelligence solutions could have a substantial social impact, by guiding a large variety of users towards achieving their desired objectives in online communication.